\documentclass[aoas,MSNbibl,nameyear,dvips]{arximspdf}
\usepackage{graphicx}

\doi{10.1214/10-AOAS398H}
\referstodoi{10.1214/10-AOAS398}
\volume{5}
\issue{1}
\pubyear{2011}
\firstpage{71}
\lastpage{75}

\begin{document}
\begin{frontmatter}

\title{Discussion of: A statistical analysis of multiple temperature
proxies: Are reconstructions of surface temperatures over the last 1000~years~reliable?}
\runtitle{Discussion}
\pdftitle{Discussion on A statistical analysis of multiple temperature proxies:
Are reconstructions of surface temperatures over the last 1000 years reliable?
by B. B. McShane and A. J. Wyner}

\begin{aug}
\author{\fnms{Lasse} \snm{Holmstr\"{o}m}\corref{}\ead[label=e1]{lasse.holmstrom@oulu.fi}}

\runauthor{L. Holmstr\"{o}m}
\pdfauthor{Lasse Holmstrom}

\affiliation{University of Oulu}

\address{Department of Mathematical Sciences\\
University of Oulu\\
POB 3000, 90014 University of Oulu\\
Finland\\
\printead{e1}} 
\end{aug}

\received{\smonth{9} \syear{2010}}
\revised{\smonth{9} \syear{2010}}



\end{frontmatter}

This is an impressive paper. The authors present a thorough examination
of the ability of various climate proxies to predict temperature. The
prediction method is one much used in climate science literature and
assumes a linear relationship between the proxies and the temperature.
The idea is to use instrumental temperature data together with the
corresponding proxy records to estimate a regression model to which
historical proxy values are then input in order to produce a backcast of
past temperature variation. The authors demonstrate convincingly that
the data used in Mann et al. (\citeyear{Mann2008}) does not allow reliable temperature
prediction using this approach and that purely random artificial proxy
records, in fact, perform equally well or even better.

While this is certainly striking and thought-provoking, one should not
be left under the impression that this is the standard approach to
understanding past climate and that temperature reconstruction per se is
impossible. In fact, some of the ``proxies'' used in the paper are
themselves supposedly successful temperature reconstructions and
therefore arguably of a more fundamental character than the predictor
produced by the Lasso. There is, for example, a long and well-established
paleoecological tradition of quantitative environmental reconstruction
based on diatoms, pollen, chironomids and other biological proxies that
in some important aspects differs from the regression approach used in
the present paper and that can offer better prediction accuracy [e.g.,
Birks (\citeyear{Birks1995}); Birks et al. (\citeyear{Birks2010})].
A typical temperature reconstruction
in this tradition uses a sediment core from a selected lake together
with training data from a number of other lakes to backcast temperatures
hundreds or thousands of years in time. As a proxy~one can use the
relative abundances of various organisms (say, different diatom taxa)
measured at various depths along the core. A model for the dependence
between the temperature and the abundances is built using a training set
that consists of the relative abundances of the same the same organisms
in surface sediment samples from a large number of lakes located in the
same general area as the core lake and their current temperatures, mean
July temperature being a typical climate variable. The training lakes
are selected to cover a wide range of environmental conditions to make
possible temperature backcasting to times when the conditions at the
core lake possibly were very different from what they are today. This is
referred to as space-for-time substitution.

Compared to the approach studied in this paper, at least two differences
seem clear: the relative locality of the analysis and the possibility in
some cases to incorporate ecological information in the model. Instead
of a global proxy network, this method typically uses local proxy data
to backcast local environmental conditions. Ecological information can
enter the model, for example, through a hierarchical model component
that explicitly incorporates the fact that different organisms have
different optimal temperatures, that is, temperatures at which they fare
particularly well. Various models, including methods based on parametric
or nonparametric regression, as well Bayesian approaches, have been
proposed. Such reconstructions appear to quite successfully capture many
large-scale Holocene climate features such as the Medieval Warm Period
and the Little Ice Age. I wonder if the rather strict locality of the
approach combined with at least some degree of ecological plausibility
in modeling the proxy data generating process are factors in their
apparent success?

Recently, Bayesian approaches have begun to enter the field. The first
papers that used detailed Bayesian modeling for paleoclimate
reconstruction were Vasko, Toivonen and Korhola (\citeyear{Vasko2000}),
Toivonen et al. (\citeyear{Toivonen2001}) and
Korhola et al. (\citeyear{Korhola2002}). More recently, Haslett et al.
(\citeyear{Haslett2006}) and
Er\"{a}st\"{o} and Holmstr\"{o}m (\citeyear{Er2006}) have taken similar approaches.
As reconstruction in the Bayesian setting produces the posterior
distribution of the past temperature variation, such important questions
as joint (pathwise) credibility of past temperature features also
discussed in the present paper can be easily answered.

\begin{figure}

\includegraphics{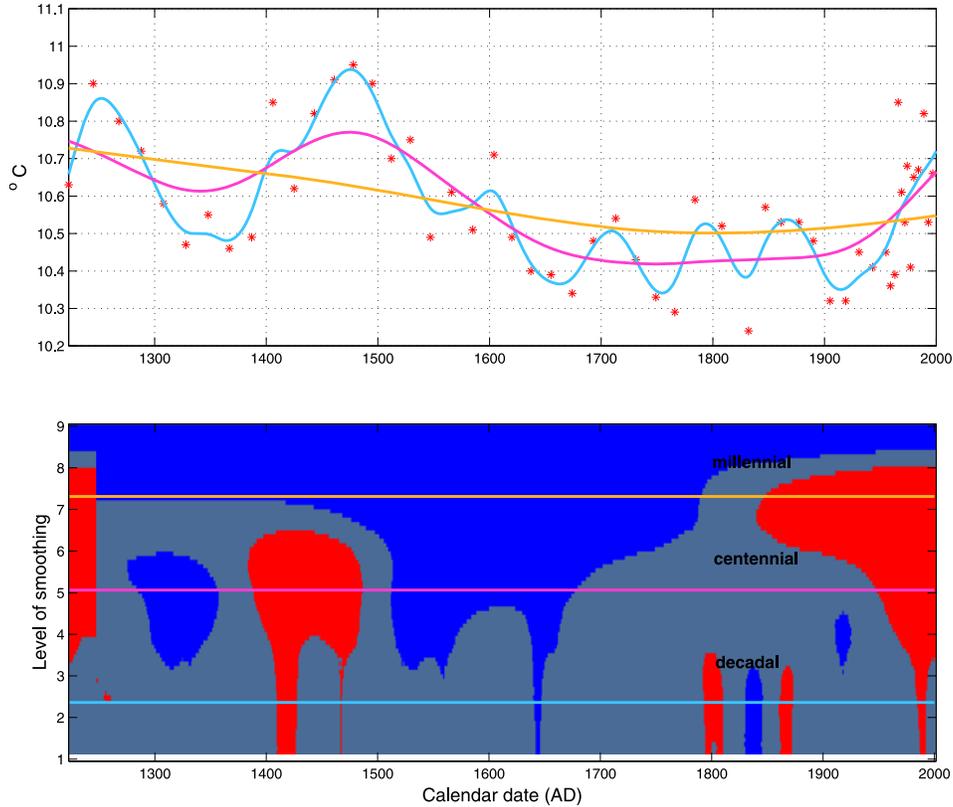}

\caption{Scale space analysis of the salient features of a diatom-based
temperature reconstruction in Northern Fennoscandia for the past 800
years. \textup{Upper panel:} the reconstructed temperatures (asterisks) together
with three smooths. \textup{Lower panel:} scale space credibility map with the
horizontal lines indicating the corresponding three levels of smoothing.
For more details, see Weckstr\"{o}m et al. (\protect\citeyear{Weckstr2006}) and the text. Code and
data to reproduce this figure can be found in the supplementary file to
this discussion.}
\label{fig1}\end{figure}

Another question I would like to consider is the possible role of
smoothing, briefly touched upon at the end of Section 3. As pointed out
by the authors, one difficulty is the choice of a proper smoothing
level. However, in the so-called scale space approach this difficulty is
turned into an opportunity when, instead of just one, in some sense
optimal smooth, one  considers a whole family of smooths. Our
paleoecologist collaborators have found quite useful~a procedure where
such multi-level smoothing is applied to the reconstructed temperature
time series. Each smooth can then be interpreted to provide information
about the underlying past temperature variation at a particular time
scale, little smoothing showing the short time scale details and heavy
smoothing leaving only the coarsest features, such as the overall trend.
For a Bayesian reconstruction, such scale space smoothing can easily be
combined with a credibility analysis of these multi-scale features, but
similar analyses are possible also in a non-Bayesian setting [e.g.,
Er\"{a}st\"{o} and Holmstr\"{o}m (\citeyear{Er2005,Er2006,Er2007}); Holmstr\"{o}m and
Er\"{a}st\"{o} (\citeyear{Holmstr2002});
Korhola et al. (\citeyear{Korhola2000});
 Godtliebsen, Olsen and Winther (\citeyear{Godtliebsen2003});
Rohling and P\"{a}like  (\citeyear{Rohling2005})].

It is also possible to combine a non-Bayesian reconstruction with
Bayesian scale space analysis. This is useful as the vast majority of
existing paleoreconstructions are non-Bayesian. Thus, suppose that $\mu (t_i)$ is
the true past \mbox{temperature at time $t_i$}, $i=1,\ldots,n$, and let $y_i=\mu(t_i)+\varepsilon_i$ be its reconstructed
value. After specifying pri- ors for $\mu$ and the reconstruction errors $\varepsilon_i$ the
posterior distribution $p(\mu'|y_1,\ldots,y_n)$ of the derivative of $\mu$ can then be obtained. The
scale space analysis now consists of smoothing this posterior at
different levels in order to find the credible features of past
temperature variation in different scales. It is also possible to handle
correlated errors as well as errors in the time points $t_i$. For details see
Er\"{a}st\"{o} and Holmstr\"{o}m (\citeyear{Er2005,Er2007}).

An example of this idea is shown in Figure~\ref{fig1}, where an analysis of a
diatom-based temperature reconstruction in Northern Fennoscandia for the
past 800 years is shown. In the upper panel, asterisks show the actual
reconstructed temperatures together with three smooths (posterior means)
reflecting variation in millennial, centennial and decadal timescales.
The lower panel map summarizes the scale space inference. Following any
horizontal line in this map we see when, at that particular smoothing
level, or time scale, the local temperature trend has been credibly
positive (red) or negative (blue) or when no credible change has
occurred (gray). Inference is joint over all time points. In this
analysis, the Little Ice Age shows as a credible broad minimum in
several different scales, from century to millennial scales (the color
changing from blue to red), whereas the overall millennial trend has
been negative. The very recent warming after industrialization shows as
red in many scales.

\begin{supplement}
\stitle{Supplement}
\slink[doi]{10.1214/10-AOAS398HSUPP}  
\slink[url]{http://lib.stat.cmu.edu/aoas/398H/supplementH.zip}
\sdatatype{.zip}
\sdescription{Code and data to reproduce Figure~\ref{fig1} can be found in the supplementary
files to this discussion.}
\end{supplement}


\printaddresses

\end{document}